\documentclass[mathleft,fleqn,
]{an}
\usepackage{graphicx}
\usepackage[varg]{txfonts}
\usepackage{natbib}
\bibpunct{(}{)}{;}{a}{}{,}
\begin{document}

\Pagespan{1}{}
\Yearpublication{2014}%
\Yearsubmission{2014}%
\Month{0}%
\Volume{999}%
\Issue{0}%
\DOI{asna.201400000}%

\title{Non-LTE iron abundances in cool stars: The role of hydrogen collisions}

\author{Rana Ezzeddine\inst{1}\fnmsep\thanks{Corresponding author:
        {rana.ezzeddine@umontpellier.fr}}
\and  Thibault Merle\inst{2}
\and Bertrand Plez\inst{1}
}
\titlerunning{Instructions for authors}
\authorrunning{T.\,H.\,E. Editor \& G.\,H. Ostwriter}
\institute{
Laboratoire Univers et Particules de Montpellier, Universit\'{e} de Montpellier, CNRS UMR-5299, Montpellier, France
\and 
Institut d'Astronomie et d'Astrophysique, Universit\'{e} Libre de Bruxelles, CP 226, Boulevard du Triomphe, 1050 Brussels, Belgium}

\received{15/09/2015}
\accepted{01/12/2015}
\publonline{XXXX}


\abstract{
In the aim of determining accurate iron abundances in stars, this work is meant to empirically calibrate H-collision 
cross-sections with iron, 
where no quantum mechanical calculations have been published yet.
Thus, a new iron model atom has been developed, which includes hydrogen collisions for excitation, ionization and charge transfer processes. We show that
collisions with hydrogen leading to charge transfer are important for an accurate non-LTE modeling. We apply our calculations on several benchmark stars
including the Sun, the metal-rich star $\alpha$~Cen~A and the metal-poor star HD140283. 
}

\maketitle

\section{Introduction}
Iron plays an important role in studying the atmospheres of cool stars. It is often used as a proxy to the total metal content and 
metallicities in stars.\\

Neutral iron lines, Fe~I, have been shown to be subject to non-LTE effects \citep{thevenin1999,mashonkina2011,lind2012}.
This deviation from LTE grows towards lower metallicities, due
to a decreasing number of electrons donated by metals which decreases the collisional rates. Thus, a non-LTE modeling of the
spectra of these stars becomes important, which in turn requires a good knowledge of a bulk of atomic data for each atom
under consideration. A common problem in non-LTE calculations comes from 
uncertainties in the underlying atomic data,
of which, in cool stars, the inelastic neutral hydrogen collisional rates are the most significant source.\\

\noindent Quantum calculations for hydrogen collisional rates have recently been calculated for a small number of elements including 
Li \citep{belyaev2003}, Na \citep{barklem2010}, Mg \citep{belyaev2012}, Al \citep{belyaev2013} and Si \citep{belyaev2014}.
For iron, however, no quantum calculations have been published yet.\\
In the lack of quantum data, a common practice is to estimate the hydrogen collisional rates using the classical Drawin approximation 
\citep{drawin1968,drawin1969a} which is a modified version of \citet{thomson1912} classical e$^{-}$ + atom ionization rate equation,
extended by Drawin to that of same atoms $(A \, + \, A)$ excitation collisions, where $A$ corresponds to an element species.\\
\noindent Drawin's approximation was then rewritten by \citet{steenbock1984} for inelastic H + atom collisions,
which was then reviewed and re-derived by \citet{lambert1993}. Both their approaches
apply only to allowed excitation collisional transitions due to the dependence of the collisional rates on the transition's
oscillator strength $f$-value in Drawin's equation.\\ 

Upon comparison with quantum calculations for Li, Na and Mg, the Drawin approximation has been shown to overestimate the collisional 
rates by several orders of magnitude \citep{barklem2011},
which is commonly treated by applying a multiplicative scaling fudge factor S$_{\mathrm{H}}$ to the Drawin rates by using
different calibration methods on reference stars.  

Recent non-LTE abundance studies using quantum calculations revealed that the charge transfer (CT) process, 
i.e. $A \, + \, \mathrm{H} \rightleftharpoons A^{+} \, + \, \mathrm{H}^{-}$, can play a more important
role than excitation (i.e. bound-bound) transitions \citep{osorio2015,lind2011}. To our knowledge, no study has yet
tested the inclusion of H+Fe charge transfer collision in their non-LTE calculations, which was an important reason that motivated this work. \\

In this article, we aim at testing the role of the different H+Fe collisional processes including excitation, ionization and charge
transfer in non-LTE calculations, using a newly developed iron model atom, and starting from well defined non-spectroscopic atmospheric  
parameters for a set of benchmark stars.

\section{Method}
We performed non-LTE, 1D modeling of Fe~I and Fe~II spectral lines using the radiative transfer code \texttt{MULTI2.3} \citep{Carlsson1986,carlsson1992},
which solves the statistical equilibrium and radiative transfer equations simultaneously for the element in question, through the Accelerated
Lambda Iteration (ALI) approximation 
\citep{scharmer1981}.
In the sections below, we present the observational data including the spectra and measured equivalent widths in Sect.
\ref{obs_data}, the model atmospheres and atmospheric parameters 
adopted for the stars under study in Sect. \ref{model atm}, and the newly developed Fe~I/Fe~II model atom used in the non-LTE calculations
in Sect. \ref{model atom}.

\subsection{Observational data} \label{obs_data}
The spectra of the stars in this work were obtained from the \textit{Gaia}-ESO survey collaboration. They were observed by the UVES spectrograph
of the VLT, and reduced by the standard UVES pipeline version 3.2 \citep{ballester2000}. All the spectra have high signal to noise ratios (S/N $>$ 110),
and a high spectral resolution of $R = 70 \, 000$.\\

The equivalent widths (EW) for each star were measured automatically using a Gaussian fitting 
method with the Automated Equivalent Width Measurement code
\texttt{Robospect} \citep{robospect2013}. The Fe~I and Fe~II linelist chosen in the analysis is a subset of the
Gaia-ESO survey ``golden'' v4 linelist \citep{heiter2015b}, which was tested on the Sun. All lines that are blended and 
whose relative errors $\bigg(\frac{\sigma EW}{EW}\bigg) > 0.3$, were removed from the linelist. In addition, only lines with EW 
between 10~m\AA \, and 150~m\AA \, in the Solar spectrum were included. The final number of lines used for each star is $\sim$ 100 Fe~I lines and 10 Fe~II lines.

\subsection{Model atmospheres}\label{model atm}
Three benchmark stars of different metallicities and stellar parameters were considered in this study, namely the Sun, the metal-rich dwarf
$\alpha$~Centauri~A and the metal-poor halo subgiant HD140283.
Their atmospheric parameters were adopted from the study of the \textit{Gaia} benchmark stars by \citet{jofre2014a}, 
where the effective temperatures $T_{\mathrm{eff}}$ and surface gravities log $g$ were determined homogeneously and independently from spectroscopic models,
while the stellar metallicities were determined spectroscopically from Fe~I lines by fixing $T_{\mathrm{eff}}$ and log $g$ to the
previous values and applying a line-by-line non-LTE correction to each line. The parameters are listed in Table \ref{tab_atm_param}.\\

1D MARCS \citep{gustafsson2008} atmospheric models were
interpolated to the atmospheric parameters (in $T_{\mathrm{eff}}$, log~$g$ and [Fe/H]) of each star using the code \texttt{interpol\_marcs.f} 
written by Thomas Masseron\footnote{http://marcs.astro.uu.se/software.php}, except for the Solar case where the reference Solar MARCS model 
was used. Background line opacities except iron, calculated for each star as a function of
its atmospheric parameters and sampled to $\sim 153 \, 000$ wavelength points using the MARCS opacity package,
were also employed in the \texttt{MULTI2.3} calculations.\\

\begin{table}
\centering
\caption{Atmospheric parameters of the stars used in this study, adopted from \citet{jofre2014a}.}
\label{tab_atm_param}
\begin{tabular}{c | c c c}\hline
\hline
Star & $T_{\mathrm{eff}}$ & log $g$ & [Fe/H] \\ 
\hline
Sun & 5777 & 4.43 & +0.00 \\
$\alpha$ Cen A & 5840 & 4.31 & +0.26\\
HD140283 & 5720 & 3.67 & -2.36\\
\hline
\end{tabular}
\end{table}

\subsection{Model atom}\label{model atom}
A new iron model atom including Fe~I and Fe~II energy levels, as well as the Fe~III ground level has been developed 
with the most up-to-date atomic data available, including radiative and collisional transitions for all the levels.
 
\subsubsection{Energy levels}
The Fe~I and Fe~II energy levels were adopted from the NIST database\footnote{http://www.nist.gov/pml/data/asd.cfm} 
from the calculations of \citet{nave1994} and \citet{nave2013} respectively and supplemented by the
predicted high-lying Fe~I levels from \citet{petkur2015} up to an excitation energy of 8.392 eV. The model was completed with the ground
Fe~III energy level. In order 
to reduce the large number of energy levels (initially 1939 fine structure levels), all the levels in our iron model atom, except the
ground and first excited states of Fe~I and Fe~II, were grouped into mean term levels from their respective fine structure levels using the code \texttt{FORMATO} \normalsize 
(Merle et al., in prep.). In addition, all mean levels above
5 eV and lying within an energy interval of 0.0124 eV (100 cm$^{-1}$)
were combined into superlevels. The excitation potential of each superlevel is a weighted
mean by the statistical weights of the excitation potentials of its corresponding mean levels.\\

The final number of levels in the model atom is 135 Fe~I levels (belonging to 911 fine structure levels and 203 spectroscopic terms) 
and 127 Fe~II levels (belonging to 1027 fine structure levels and 189 terms).

\subsubsection{Radiative transitions}
For our Fe~I/Fe~II model, we used the VALD\small{3} \normalsize \citep{ryabchikova2011} interface 
database\footnote{http://vald.inasan.ru/$\sim$vald3/php/vald.php} to extract all
the Fe~I and Fe~II radiative bound-bound transitions. In addition, the UV and IR lines corresponding to transitions from and
to the predicted high lying levels have also been included in the model \citep{petkur2015}. 
Individual transitions belonging to levels that have been combined to superlevels have also
been combined into superlines using \texttt{FORMATO}. The superline total
transition probability is a weighted average of $gf$-values of individual transitions, combined via the relation of \citet{martin1988}.
Our final FeI/FeII model includes 9816 FeI (belonging to 81162 lines) and 16745 FeII (belonging to 113964 transitions)
super transitions combined from the individual lines.\\

In addition to the b-b transitions, Fe~I and Fe~II energy levels were coupled via photoionization to the Fe~II and Fe~III ground levels respectively.
For Fe~I levels, the corresponding photoionization cross-section tables were calculated by \citet{bautista1997} for 52 LS terms and those for Fe~II
by \citet{nahar1994} for 86 LS terms, and were extracted from the NORAD\footnote{http://www.astronomy.ohio-state.edu/$\sim$csur/NORAD/norad.html} 
database (Nahar \& Collaboration). For the rest of the levels, the hydrogenic approximation was used to calculate threshold cross-sections
via Kramer's semi-classical relation \citep{kramer1968}.
All the photoionization energies extracted from NORAD were shifted to match the threshold ionization energies in NIST due to existing energy differences 
between their theoretical values \citep{verner1994}. Sharp cross-section resonance peaks in the tables were smoothed as a function of photon frequencies and then using an opacity sampling method,
they were resampled to a maximum of 200 frequency points per transition.

\subsubsection{Collisional transitions} \label{col_trans}
All levels in our iron model were coupled via electron and neutral hydrogen atom collisional transitions. e$^{-}$ b-b effective collisional strengths 
were included from the calculations of \citet{pelan1997} for the ground and first excited states of Fe~I, and 
from \citet{zangprad1995} and \citet{bautista1996} for 142 Fe~II
fine structure levels. For the rest of the levels, the \citet{seaton1962a} and \citet{seaton1962b} impact approximations were used to calculate 
the cross-sections for the allowed and forbidden transitions respectively. In addition, e$^{-}$ ionization collisional transitions were included using the 
semi-classical approximation of \citet{bely1970}.\\

For the H collisions, \citeauthor{lambert1993}'s (\citeyear{lambert1993}) derivation of the Drawin approximation was used to calculate the rate coefficients $\langle \sigma \nu \rangle$, 
with the oscillator strength $f$ set to 1 for all transitions. This was motivated by the approach adopted by \citet{steenbock1985} who set the 
$Q$-factor\footnote{$Q = \bigg(\frac{\chi^{\infty}_{\mathrm{H}}}{\Delta E^{\mathrm{A}}}\bigg)^{2} f$, 
where $\chi^{\infty}_{\mathrm{H}}$ = 13.6 eV is the ionization potential of hydrogen, $\Delta E^{\mathrm{A}}$ is the transition energy of atom $A$ and 
$f$ is the oscillator strength of the transition.} in their Mg+H 
rate equations equal to 1 for forbidden transitions. Another attempt to remove the oscillator strength dependence from the rate equations was by \citet{collet2005}, who set 
a constant $f=10^{-3}$ in the \citet{vanregemorter1962} equation for the e$^{-}$ collisions for all transitions. In addition, recent quantum calculations for Mg and other elements
found large H-collisional rates for forbidden transitions, which were comparable to the allowed ones \citep{feautrier2014}.\\
We also adopt the Drawin approximation for charge transfer H collisions in the absence of any other suitable approximation.\\

All the levels in the model were coupled with b-b, b-f and charge transfer
($\mathrm{Fe} + \mathrm{H} \, \, \, \rightleftharpoons \, \, \, \mathrm{Fe^{+}} + \mathrm{H^{-}} $) H collisions. 
The rates were scaled with a different global multiplicative scaling factor, as follows:
\begin{itemize}
 \item S$_{\mathrm{H}}$: multiplicative factor for b-b and b-f H collision rates.
 \item S$_{\mathrm{H}}$(CT): multiplicative factor for charge transfer H collision rates.
\end{itemize}

Several modified model atoms were created using different scaling factors for each H collisional process. S$_{\mathrm{H}}$ was varied 
between 0.0001 and 10, in multiplication steps of 10, while S$_{\mathrm{H}}$(CT) 
was varied between 0.1 and 10 in multiplication steps of 10, in addition to the S$_{\mathrm{H}}$(CT)=0 models. 

\section{non-LTE calculations and results}
\begin{figure*}
\includegraphics[width=\textwidth,height=80mm]{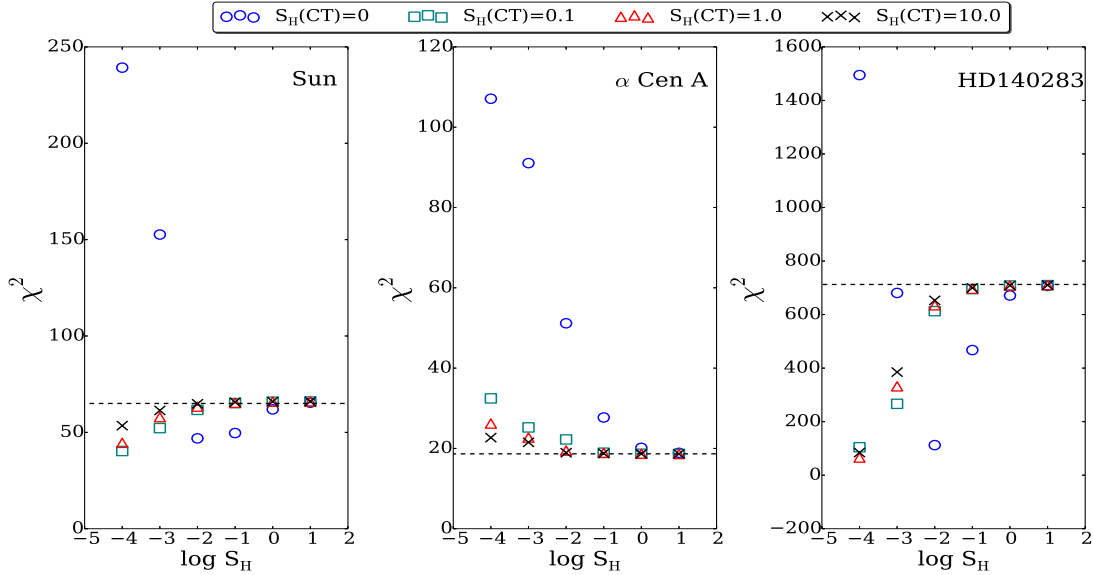}
\caption{$\chi^{2}$ as a function of scaling factors S$_{\mathrm{H}}$ and S$_{\mathrm{H}}$(CT) for the Sun, HD140283 
and $\alpha$ Cen A. The dotted lines represent the $\chi^{2}$ values obtained in LTE.}
\label{fig_chisq_stars}
\end{figure*}

Non-LTE calculations were performed for each atomic model with a given S$_{\mathrm{H}}$ and S$_{\mathrm{H}}$(CT).
Thus, a total of 24 models were computed for each star.
The calculated EW(calc) for each model were compared to the measured EW(obs) using the $\chi^{2}$ test:
 \begin{equation}
 \chi^{2} = \frac{1}{\mathrm{N_{\mathrm{lines}}}} \sum_{\mathrm{lines}} \bigg(\frac{\mathrm{EW}(\mathrm{calc}) - 
 \mathrm{EW}(\mathrm{obs})}{\sigma \mathrm{EW}(\mathrm{obs})}\bigg)^{2}
\end{equation}
\noindent where N$_{\mathrm{lines}}$ is the total number of lines used in the $\chi^{2}$ test. The variation of $\chi^{2}$ as a function of 
S$_{\mathrm{H}}$ and S$_{\mathrm{H}}$(CT) for each star is shown in Fig. \ref{fig_chisq_stars}. The LTE $\chi^{2}$ are also shown for
comparison.\\

For the Sun, it can be seen that when charge transfer rates are neglected, the best fit is obtained at S$_{\mathrm{H}}$=0.01. Upon including
charge transfer rates, smaller S$_{\mathrm{H}}$ values are needed. For the metal-rich star $\alpha$~Cen~A, larger values of
S$_{\mathrm{H}}$ and S$_{\mathrm{H}}$(CT) are favored.
For the metal-poor star HD140283, large variations in $\chi^{2}$ are obtained with S$_{\mathrm{H}}$ and S$_{\mathrm{H}}$(CT) showing that
charge transfer process plays an important role in producing the best-fit. Similar to the Sun,  
the best fit is also obtained at S$_{\mathrm{H}}$=0.01 upon neglecting the charge transfer rates, while smaller values of S$_{\mathrm{H}}$
are needed when including them.\\

Comparing to recent studies,
\citet{mashonkina2011} and \citet{bergemann2012} used Fe~I/Fe~II model atoms in their non-LTE abundance determinations,
which are comparable to our model. Their calculations were also tested on benchmark stars including the Sun and HD140283,
where they used the Drawin approximation scaled with an S$_{\mathrm{H}}$-factor for the excitation and ionization H collisional rates.
Charge transfer rates were not included in their calculations.
They could not find a single S$_{\mathrm{H}}$-factor that would fit all stars.
For metal-poor stars, \citet{mashonkina2011} determined a value of S$_{\mathrm{H}}$=0.1 while \citet{bergemann2012} determined an optimum value of
S$_{\mathrm{H}}$=1. For solar-metallicity stars, both studies found that different S$_{\mathrm{H}}$ values had no significant effect on the
calculated abundances.\\

Similarly, we could not find a single set of S$_{\mathrm{H}}$ values that would ensure a best fit for all stars. We could, however, note that 
including charge transfer H collisional
rates is important in iron non-LTE calculations. When neglecting charge transfer rates, however, a large value of S$_{\mathrm{H}}$ is needed for 
$\alpha$~Cen~A (S$_{\mathrm{H}}$ $>$ 1), while a smaller value is needed for the Sun and HD140283 (S$_{\mathrm{H}}$ $\leq$ 0.01),
 which is not in accordance with previous studies.

\section{Conclusions}
We performed iron non-LTE spectral line calculations for three benchmark stars with well determined atmospheric parameters, using
hydrogen collisions for excitation and ionization processes, and including charge transfer rates for the first time. 
We show that the charge transfer rates
are important to include in the non-LTE calculations, especially for the metal-poor star.
They were found, however, to play a less important role with increasing metallicity 
for the Sun and the metal-rich star $\alpha$~Cen~A, where non-LTE effects are of smaller magnitude.
No single set of values for the scaling factors S$_{\mathrm{H}}$ and S$_{\mathrm{H}}$(CT) was obtained for the different types of stars.
This demonstrates the inability of the Drawin approximation to
reproduce the correct behavior and magnitudes of hydrogen collision rates (see \citet{barklem2011}).
In the lack of quantum calculations for the hydrogen collision rates, more efficient models than the classical Drawin 
approximation are required. We are working on such a method based on semi-empirical fitting of the available quantum data for other chemical species.

\acknowledgements We would like to acknowledge the GES-CoRoT collaboration for providing us with the UVES spectra for the benchmark stars,
which have been used in this work. This work has made use of the VALD database, operated at Uppsala University,
the Institute of Astronomy RAS in Moscow, and the University of Vienna.


%
%

\end{document}